\documentclass{ifacconf}

\usepackage{graphicx}      
\usepackage{amsmath}
\usepackage{amssymb}
\usepackage{natbib}        
\usepackage[usenames,dvipsnames]{xcolor}

\begin{document}
\begin{frontmatter}

\title{Aperiodic Communication for MPC in Autonomous Cooperative Landing\thanksref{footnoteinfo}} 

\thanks[footnoteinfo]{This work was supported by the Wallenberg AI, Autonomous Systems and Software Program (WASP) and the Swedish Research Council, Knut and Alice Wallenberg Foundation (KAW).}

\author[First]{Dženan Lapandić}, 
\author[First]{Linnea Persson},
\author[First]{Dimos V. Dimarogonas}, 
\author[First]{Bo Wahlberg}

\address[First]{Division of Decision and Control Systems, School of Electrical Engineering and Computer Science, KTH Royal Institute of Technology, Stockholm, Sweden\\
(e-mail: lapandic, laperss, dimos, bo@kth.se).}

\begin{abstract}                
This paper investigates the rendezvous problem for the autonomous cooperative landing of an unmanned aerial vehicle (UAV) on an unmanned surface vehicle (USV). Such heterogeneous agents, with nonlinear dynamics, are dynamically decoupled but share a common cooperative rendezvous task. The underlying control scheme is based on distributed Model Predictive Control (MPC). The main contribution is a rendezvous algorithm with an online update rule of the rendezvous location. The algorithm only requires the agents to exchange information when they can not guarantee to rendezvous. Hence, the exchange of information occurs aperiodically, which reduces the necessary communication between the agents. Furthermore, we prove that the algorithm guarantees recursive feasibility. The simulation results illustrate the effectiveness of the proposed algorithm applied to the problem of autonomous cooperative landing.
\end{abstract}


\end{frontmatter}

\section{Introduction}

Coordination and control of multi-agent systems is a vivid research area with applications in robot manipulators control, unmanned surface vehicles (USV), unmanned aerial vehicles (UAV) and space systems, among others. 
Because multi-agent systems are composed of agents with embedded computing and communication units, a distributed control scheme is the most common control approach to these types of problems. 

\begin{figure}[tb]
\begin{center}
\includegraphics[width=7.0cm]{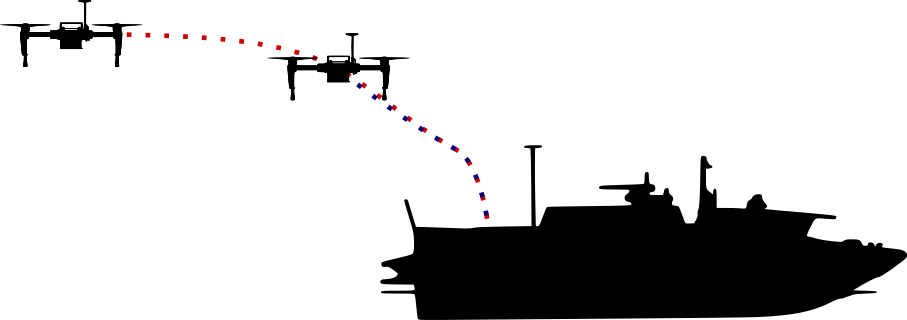}    
\caption{The motivating application is a scenario where drones must be able to rendezvous and land on a moving boat.}
\label{fig:ilustration}
\end{center}
\end{figure}

Search-and-rescue missions are one example of an application that is dependent on distributed and multi-agent control. In such a mission, heterogeneous agents have to perform tasks together or independently while considering the common objective of the mission and assisting other agents if needed. 
This type of scenario has been tested as a part of the WASP Research Arena on Public Safety, \cite{persson2019model}.
The problem of safely landing UAVs on USVs while they are moving at high speeds to ensure agents rendezvous simultaneously has been studied in \cite{perssonBW}. The rendezvous problem is challenging due to several reasons, for example, sudden communication losses or strong disturbances acting on the agents can lead to disastrous consequences. Moreover, even the basic tasks to determine if the rendezvous is possible or not and what strategy to employ when the rendezvous location has to be updated can be complex. An illustration of the motivating problem is depicted in Fig.~\ref{fig:ilustration}.

Model Predictive Control (MPC) has often been used in such applications because of its ability to explicitly include advanced system dynamics as well as diverse state and input  constraints directly in the computation of the control inputs.
A question that has not been directly addressed in previous research is that of efficient communication strategies between the agents. 
Instead, previous distributed solutions have exchanged all state and trajectory information between the agents at each sample time,~\cite{berezapersson}. 
In this paper, we consider rendezvous control through Distributed MPC (DMPC), where the agents use an  aperiodic exchange of information
to negotiate and update their rendezvous point. 
The agents achieve cooperation through the iterative updates of the shared rendezvous point. 
The exchange of information occurs only when it is necessary to maintain the feasibility of the control action, thus reducing the necessary communication between the agents.
The control algorithm is applied to  nonlinear heterogeneous agents with state and input constraints, and tested and evaluated in simulation on an example of a UAV landing on a USV.

The main contributions of this paper are outlined as follows: 
\begin{itemize}
\item We present the distributed rendezvous algorithm that enables the aperiodic communication between the agents based on the deviations from the predicted trajectory, thus eliminating unnecessary communication.
\item Moreover, we synthesize the time-varying distributed terminal sets for tracking that depend on the rendezvous point. These terminal sets are the main ingredient in the recursive feasibility proof. 
\item Finally, we prove that the proposed algorithm guarantees recursive feasibility.
\end{itemize}

\cite{christofides2013distributed} gives an overview of several approaches to distributed implementation of model predictive control. Our focus is on dynamically decoupled systems that can be coupled with performance criteria. In \cite{keviczky2006decentralized}, the authors assume that each agent knows the system dynamics of all of its neighbors to compute their assumed optimal state trajectories. The stability is established with the requirement that the mismatch from the actual trajectories of the agent's neighbors is small. A similar approach was taken in \cite{dunbar2006distributed}, in which the stability is imposed by requiring that the calculated trajectories of each agent do not deviate from those calculated in the previous time step. Sequential optimization of the local cost functions can, under some assumptions, guarantee stability and convergence to the common cooperative goal, as shown in \cite{muller2012cooperative}.
In our approach, we are considering the agents that are unaware of the dynamics of other agents and achieve the cooperative goal by negotiating the rendezvous location.

However, most of the mentioned research assumes a periodical exchange of information between the agents and recalculation of the control inputs at every sampling time instance. The recalculated control inputs usually do not generate much different state trajectories compared to the ones from the previous time steps, especially if the model is very accurate and disturbances acting on the system are small but are critical for feasibility requirements, see, e.g. \cite{chen1998quasi}. 
The aperiodic (distributed) MPC can be implemented using the event-triggered or self-triggered strategy \cite{heemels2012introduction}. The triggering conditions can be cost-based, then the optimal control problem is recalculated when the cost is not guaranteed to decrease \cite{hashimoto2014distributed}. Moreover, they can be trajectory-based and recalculated when the trajectories deviated significantly compared to the previous ones and the feasibility of the overall problem might be compromised \cite{hashimoto2017event}, \cite{liu2020distributed}. However, the triggering conditions for nonlinear systems are based on the worst-case trajectory prediction that involves Lipschitz continuity assumption and Lipschitz constant, which for the systems with fast and agile dynamics, like quadcopters, can lead to very conservative triggering conditions to maintain feasibility and stability. Therefore, in this paper, we assume that the recalculation of the optimal control problem is conducted at every time step, and investigate how aperiodic negotiation of the rendezvous location can preserve the feasibility.

The paper is organized as follows. First, we state the problem formulation and the distributed optimal control problem in Section~\ref{sec:problem_formulation}. Then, we present the rendezvous algorithm in Section~\ref{sec:rendezvous_algorithm_sec} and its feasibility in Section~\ref{sec:feasibility_sec}. Finally, in Section~\ref{sec:simulation_results}, we describe the models and their constraints used to generate the results that are also presented in this section.

\textbf{Notation:} We use $P\succ 0$ to denote that a matrix $P$ is positive definite. The notation $\left \| x \right \|$ is used as the Euclidean norm of vector $x$, and $\left \| x \right \|_P$ as a weighted norm of $x$, where $\left \| x \right \|_P = \sqrt{x^TPx}$. 
We denote the system state trajectories with $x(t)$, nominal state trajectories with $\hat{x}(t)$ and optimal state trajectories with $\hat{x}^*(t)$. 

\section{Problem Formulation}\label{sec:problem_formulation}
\subsection{Dynamics and optimal control problem}
We consider $M$ agents with nonlinear dynamics and additive disturbances:
\begin{align}
\begin{split}
        \dot{x}_i(t)&=f_i(x_i(t),u_i(t)) + w_i(t), \\
    y_i(t)&=C_ix_i(t), \label{eq:1}
\end{split}
\end{align}
for $t\geq t_0$, where for each $i=1,...,M$,  the state vector $x_i(t)\in \mathbb{R}^{n_i}$ is measurable, $u_i(t)\in \mathcal{U} \subseteq  \mathbb{R}^{m_i}$ is the control input, the output $y_i(t)\in \mathbb{R}^{p}$ consists of the states we aim to control for the rendezvous, $w_i(t)  \in \mathcal{W} \subseteq \mathbb{R}^{n_i}$ is the additive bounded disturbance, and $t_0 \in \mathbb{R}$ is the initial time.

The following standard MPC assumptions as in \cite{chen1998quasi} are considered in this paper.
\begin{assum}\label{as:standard}
(i) The function $f_i:\mathbb{R}^{n_i}\times \mathbb{R}^{m_i}\rightarrow \mathbb{R}^{n_i}$ is twice continuously differentiable and $f_i(0,0)=0$; 
(ii) $\mathcal{U} \subseteq  \mathbb{R}^{m_i}$ is compact, convex and $0\in  \mathbb{R}^{m_i}$ is contained in $\mathcal{U}$; 
(iii) the system in (\ref{eq:1}) has a unique solution for any initial condition $x_{i,0} \in \mathbb{R}^{n_i}$, any piecewise continuous and right-continuous control $u_i:\left [ t_0, \infty \right ) \rightarrow \mathcal{U}_i$, and any disturbance $w_i:\left [ t_0, \infty \right ) \rightarrow \mathcal{W}_i$; (iv) for the linearized system around the origin without disturbances, i.e., $\dot{x}_i=A_ix_i(t)+B_iu_i(t)$, where $A_i=\frac{\partial f_i}{\partial x_i}(0,0)$ and $B_i=\frac{\partial f_i}{\partial u_i}(0,0)$, the pair $(A_i,B_i)$ is stabilizable; (v) for each agent $i$ and its linearized dynamics around the origin, there exists a matrix $K_i$ such that $A_{k,i}=A_i+B_iK_i$ is a stable Hurwitz matrix.
\end{assum}

\begin{rem}
Note that the requirement $f_i(0,0)=0$ is not restricted to the origin, but can be shifted to any equilibrium $(\bar{x}_{i},\bar{u}_{i})$, as well as the linearization in (iv).
\end{rem}



Let $\hat{x}_i(s;t_k),\hat{y}_i(s;t_k)$ be the nominal state trajectory and output, respectively, calculated at time instant $t_k$ given by 
\begin{align}
\begin{split}
    \dot{\hat{x}}_i(s;t_k)&=f_i(\hat{x}_i(s;t_k),u_i(s;t_k)),\label{eq:nominal} \\
    \hat{y}_i(s;t_k)&=C_i\hat{x}_i(s;t_k),
\end{split}
\end{align}
for $s \in \left [ t_k,t_k+T\right ]$.

The control objective is to steer the relevant states of every agent $y_i$ to a rendezvous point $\theta \in \mathbb{R}^p$ in finite time. The set of all admissible rendezvous points is denoted with $\Theta \subseteq \mathbb{R}^p$.
\newpage
Let us define a set $\mathcal{Z}_{i}(\theta)$ for each agent $i$ and argument $\theta \in \mathbb{R}^p$ with a tuple $(\bar{x}_i,\bar{u}_i,\bar{y}_i)$ such that $
    \mathcal{Z}_{i}(\theta) = \{ (\bar{x}_i,\bar{u}_i,\bar{y}_i) \in \mathbb{R}^{n_i+m_i+p} : 0=f_i(\bar{x}_i,\bar{u}_i), \bar{y}_i=C_i\bar{x}_i= \theta \}$
    
\begin{assum}\label{as:output_space}
There exists a non-empty compact and convex set $\Theta \subseteq \mathbb{R}^{p}$ such that $\forall \theta \in \Theta$, we have $\mathcal{Z}_{i}(\theta)\neq \emptyset$ for all $i$.
\end{assum}
Considering the motivating application, one can think of the set $\Theta$ as an inflated convex set in the plane of the USV landing platform that covers the unoccupied space that UAV and USV can reach.

By this assumption, it is also assumed that there exists an equilibrium for which the output reference $\theta$ is attained for each agent. Moreover, such an equilibrium can be explicitly found with a given $\theta$ by the following linear mappings $H_{x_i}\in \mathbb{R}^{p\times n_i}$, $H_{u_i}\in \mathbb{R}^{p\times m_i}$
\begin{equation}\label{eq:steady_states_alpha}
    \bar{x}_i = H_{x_i}\theta, \quad \bar{u}_i = H_{u_i}\theta.
\end{equation}
The following assumption is made to ensure that a such rendezvous point is reachable (in a similar manner to \textit{Assumption 2.} in \cite{keviczky2008study}):
\begin{assum}
The time planning horizon $T$ is long enough to reach at least one $\theta$ in the rendezvous set $\Theta$.
\end{assum}

We choose the cost function to penalize the deviations of the system trajectories from the desired terminal steady-state $(\bar{x}_i,\bar{u}_i,\bar{y}_i)$:
\begin{align}\label{eq:cost}
\begin{split}
    &J_i(\hat{x}_i(t_k),u_i(t_k),\bar{x}_i,\bar{u}_i) = \left \| \hat{x}_i(t_k+T;t_k) - \bar{x}_i \right \|_{P_i}^2 \\
    &+ \int_{t_k}^{t_k+T}  \left \| \hat{x}_i(s;t_k) - \bar{x}_i \right \|_{Q_i}^2 + \left \| u_i(s;t_k) - \bar{u}_i \right \|_{R_i}^2 ds, 
\end{split}
\end{align}
where $Q_i,R_i,P_i$ are positive definite weighting matrices, $T>0$ is the time duration of prediction horizon.

Note that this formulation is a bit different from the standard tracking MPC formulations (see e.g. \cite{limon2008mpc}), because of Assumption~\ref{as:output_space} that such a tuple $(\bar{x}_i,\bar{u}_i,\bar{y}_i)$ exists and is attainable.

Before we formulate the distributed optimal control problem we will present a Lemma on the local invariant terminal sets around a steady-state that is formulated following the ideas of \cite{chen1998quasi}, \cite{dunbar2007distributed}, \cite{hashimoto2017event}.
\begin{lem}
\label{lem:terminal}
For the nominal system \eqref{eq:nominal}, if Assumption~\ref{as:standard} holds, then there exists a positive constant $\alpha_i \in (0,\bar{\alpha}_i]$, a matrix $P_i=P_i^T\succ 0$, and a local state feedback control law $\kappa_{f_i}(x_i,\bar{x}_i)=K_i(x_i-\bar{x}_i) \in \mathcal{U}_i$ for a steady-state $\bar{x}_i$, satisfying
\begin{equation*}
    \frac{\partial V_{f,i}}{\partial x_i}f_i(x_i-\bar{x}_i,\kappa_{f_i}(x_i,\bar{x}_i)) \leq -\frac{1}{2}\left \| x_i-\bar{x}_i \right \|_{Q_i^*}^2
\end{equation*}
for all $x_i \in \mathcal{X}_{f,i}(\bar{x}_i,\alpha_i)$, where $Q_i^* = Q_i+K_i^T R_i K_i$, \\
$V_{f,i}(x_i,\bar{x}_i)=\left \| x_i -\bar{x}_i\right \|_{P_i}^2$ and the terminal set \begin{equation}\mathcal{X}_{f,i}(\bar{x}_i,\alpha_i) = \left \{ x_i \in \mathbb{R}^{n_i} : V_{f,i}( x_i,\bar{x}_i) \leq \alpha_i^2 \right \}.
\end{equation}
\end{lem}


%
The proof is omitted for brevity and the main parts can be found in the aforementioned papers. 

Now, we can formulate the distributed optimal control problem with respect to our objective.

\begin{prob}\label{prob:docp}
At time $t_k$ with initial states $x_i(t_k)$, $i=1,...,M$, and given reference $\theta(t_k)$, the distributed optimal control problem is formulated as 
\begin{subequations}\label{eq:mpc_problem}
\begin{equation}
    \min_{u_i(\cdot),\bar{x}_i,\bar{u}_i}J_i(\hat{x}_i(s;t_k),u_i(\cdot),\bar{x}_i,\bar{u}_i) \label{eq:8} 
\end{equation}
subject to
\begin{align}
    &\dot{\hat{x}}_i(s;t_k)=f_i(\hat{x}_i(s;t_k),u_i(s;t_k)),\enspace s \in \left [ t_k,t_k+T \right ], \label{eq:9} \\
    &\hat{y}_i(s;t_k) = C_i\hat{x}_i(s;t_k), \label{eq:10}\\
    &\hat{x}_i(s;t_k) \in \mathcal{X}_i,\label{eq:11} \\
    &u_i(s;t_k) \in \mathcal{U}_i,\label{eq:12} \\
    &\bar{x}_i = H_{x_i}\theta(t_k),\\
    &\bar{u}_i = H_{u_i}\theta(t_k),\\
    &\hat{x}_i(t_k+T;t_k)\in \mathcal{X}_{f,i}(\bar{x}_i,\alpha_i), \label{eq:13}
\end{align}\label{eq:prob}
\end{subequations}
for agents $i\!=\!1,...,M$. For the initial time $t_0$, $k\!=\!0$, the agents minimize the cost \eqref{eq:8} subject to (\ref{eq:9}--h) for a given $T>0$.
\end{prob}


\section{Rendezvous algorithm}\label{sec:rendezvous_algorithm_sec}
The distributed optimal control problem stated in~\eqref{eq:mpc_problem} depends on $\theta(t_k)$ which is the rendezvous point in the subset of the output space $\mathbb{R}^p$ as stated in \textit{Assumption~\ref{as:output_space}.} Before we present the algorithm, we need to define how $\theta(t_k)$ is going to be initialized and updated.

The rendezvous point $\theta(t_k)$ at $k=0$ can be initialized as a weighted average of the initial agent positions in the output space
\begin{equation}
    \theta(t_0) = \frac{1}{M}\sum_{i=1}^M c_i y_i(t_0), \text{ s.t. } \frac{1}{M}\sum_{i=1}^M c_i = 1, c_i\geq0, \label{eq:theta_init}
\end{equation}
where $M$ is the number of agents.

We assume that there exists $c_i$, $i=1,...,M$ such that $\theta(t_0)\in \Theta$ according to \textit{Assumption~\ref{as:output_space}.} If the agents are operating in an unconstrained and obstacle-free output space, then any $c_i$ will result with $\theta(t_0)\in \Theta$. If this is not the case, then an admissible $c_i$ would need to be determined by another layer of the optimization taking into account output-space constraints of all agents. Moreover, future work will include the conditions such that $\theta(t_k)$ remains in a constrained output space $\Theta$.

Let us denote the output terminal offset term $V_o$ as 
\begin{align}
    \begin{split}
        V_o = V_o(\hat{y}_i,\theta) &= V_o(\hat{y}_i(t_k+T;t_k),\theta(t_k)) \\
        &= \left \| \hat{y}_i(t_k+T;t_k)-\theta(t_k) \right \|^2.
    \end{split}
\end{align}

After the initialization, the agent $i$ updates $\theta(t_k)$ according to the rule 
\begin{equation}
    \theta(t_{k+1}) = \left\{\begin{matrix}
    \theta(t_k) &  V_o \leq \varepsilon \\
    \theta(t_k) - \eta v_\theta(t_k) & V_o > \varepsilon 
\end{matrix}\right. \label{eq:theta_update}
\end{equation}
where $\eta$ and $\varepsilon$ are tuning parameters and $v_\theta(t_k)$ is defined as:
\begin{equation}
    v_\theta(t_k) = \frac{\partial V_o}{\partial \theta(t_k)} \left \| \frac{\partial V_o}{\partial \theta(t_k)} \right \|^{-1}.
\end{equation}
Parameter $\eta$ is a step size that must be chosen as a small value, in order to avoid overshooting, and it quantifies the correction of $\theta$ in the output space. 

\begin{alg}(Event-triggered DMPC Rendezvous)\label{alg:et_dmpc_rendezvous}
\begin{enumerate}
     \item Initialization: Set prediction horizon $T$; sampling period $\delta$; weighting matrices $Q_i,R_i,P_i$; initial state $x_{i,0}$ at time $t_0$ for each agent $i=1,...,M$; $k=0$; $c_i$, $\theta(t_0)$ according to \eqref{eq:theta_init} and parameters $\eta$ and $\varepsilon$;\\
     \item For each agent $i=1,...,M$: 
     \begin{enumerate}
         \item If new data message received: download $\theta(t_k)$; 
         \item Solve optimization problem \eqref{eq:prob}; obtain the input $\hat{u}_i^*$; generate predicted optimal output trajectories $\hat{y}_i^*(s;t_k)$.
         \item Check the rendezvous condition:
         \begin{equation}
             V_o(\hat{y}_i(t_k+T;t_k),\theta(t_k))\leq \varepsilon \label{eq:rc}
         \end{equation}
         \begin{enumerate}
             \item If (\ref{eq:rc}) is not satisfied: update $\theta(t_{k+1})$ according to the rule \eqref{eq:theta_update}; 
              send data message $\{\theta(t_{k+1})\}$ to other agents.
         \end{enumerate}
         \item Check the stopping condition:  
         \begin{equation}
             \left \| y_i(t_k) - \theta(t_k) \right \| \leq \varepsilon
         \end{equation} 
            If not satisfied: apply $\hat{u}_i^*(t_k;t_k)$, set $k = k+1$, go to step (2)
     \end{enumerate}
     \item End
\end{enumerate}
\end{alg}

\begin{rem}
If the rendezvous condition is not satisfied, the only information that is sent from an agent $i$ at time $t_k$ is $\theta(t_k)$, and other agents use that $\theta$ as they receive it.  Therefore, the algorithm is able to run in parallel and sequentially, see e.g. \cite{richards2007robust}.
\end{rem}

\section{Feasibility}\label{sec:feasibility_sec}
In order to show feasibility of Problem~\ref{prob:docp}, we will assume the initial feasibility and then show that the problem is recursively feasible.

\begin{assum}\label{as:initial_feas}
Problem~\ref{prob:docp} is feasible at time $t_0$ for each agent $i=1,...,M$ with $\theta(t_0)$ initialized as in \eqref{eq:theta_init}. 
\end{assum}

The main point in the proof of the rendezvous algorithm is to ensure feasibility on the consecutive steps where the rendezvous reference point $\theta(t_k)$ is updated. The space shift of the terminal set $\mathcal{X}_{f,i}(\bar{x}_i,\alpha_i)$ that occurs due to the reference change $\theta(t_{k+1})\neq\theta(t_k)$ at some $t_k$ can be quantified using the update rule \eqref{eq:theta_update}.

\begin{lem}\label{lem:moving_terminal}
For the nominal system with dynamics in Eq. \eqref{eq:nominal} and reference change from $\bar{x}_i(t_k)$ to $\bar{x}_i(t_{k+1})$, given a local terminal set 
\begin{equation*}
\mathcal{X}_{f,i}(\bar{x}_i,\alpha_i) = \left \{ x_i \in \mathbb{R}^{n_i}: V_{f,i}(x_i, \bar{x}_i) \leq \alpha_i^2 \right \}
\end{equation*}
it holds that if
\begin{equation*}
    \hat{x}_i(t_k+T;t_k) \in \mathcal{X}_{f,i}(\bar{x}_i(t_k),\alpha_i(t_k))
\end{equation*}
then
\begin{equation*}
    \hat{x}_i(t_{k+1}+T;t_{k+1}) \in \mathcal{X}_{f,i}(\bar{x}_i(t_{k+1}),\alpha_i(t_{k+1}))
\end{equation*}
where $\alpha_i(t_{k+1})=\alpha_i(t_k) + \eta\left \| H_{x_i}v_\theta(t_k) \right \|_{P_i}$.
\end{lem}
The proof can be found in Appendix~\ref{ap:appA}. Now, we can state the recursive feasibility theorem.

\begin{thm}\label{thm:recursive_feasibility}
For the agents $i=1,...,M$ with system dynamics given by~\eqref{eq:1}, for which Assumptions \ref{as:standard} and \ref{as:initial_feas} and Lemmas \ref{lem:terminal} and \ref{lem:moving_terminal} hold, Problem~\ref{prob:docp} is feasible at $t_k,k\geq 0$.
\end{thm}
The proof can be found in Appendix~\ref{ap:appB}. Note that this only guarantees feasibility and does not imply convergence, which will be the focus of future work.





\section{Simulation Setup and Results}\label{sec:simulation_results}

In this section we evaluate Algorithm~\ref{alg:et_dmpc_rendezvous} implemented on nonlinear models of a quadcopter and a boat. The goal is to land the quadcopter on a boat landing platform, which is $1\text{m} \times 1\text{m}$ in size. We denote the quadcopter and the boat model and parameters with the subscripts $i=q$ and $i=b$, respectively.  
\subsection{Models and constraints}
The state vector of quadcopter model $x_q$ is chosen as
\begin{equation*}
    x_q = \left [ p_x,\ p_y,\ p_z,\ v_x,\ v_y,\ v_z,\ \phi,\ \theta,\ \psi \right ]^T,
\end{equation*}
and input $u_q$ as $u_q = \left [ \dot{v}_{z,cmd},\ \phi_{cmd},\ \theta_{cmd},\ \dot{\psi}_{cmd} \right ]^T$.

The position in $\mathbb{R}^3$ is represented with 
$y_q = \left [ p_x,\ p_y,\ p_z \right ]^T$, and $\left [ \dot{p}_x,\ \dot{p}_y,\ \dot{p}_z \right ]^T = \left [ v_x,\ v_y,\ v_z \right ]^T$. Thus, matrix $C_q=[I_{3\times3},0_{3\times6}]$.

For the derivation of the quadcopter dynamics the reader is referred to \cite{persson2019model}. The main difference is that the attitude dynamics are approximated by the inner-loop attitude dynamics that are of first order, and for the yaw angular velocity we assume that it can be instantaneously achieved, see e.g. \cite{kamel2017linear}. 



On the quadcopter we imposed several constraints to ensure the proper behaviour:
\begin{equation*}
\begin{aligned}
  \sqrt{v_x^2+v_y^2+v_z^2} &\leq 17.0 \text{ m/s}, \\
  | v_z | &\leq 4.0 \text{ m/s}, \\
  | \phi | &\leq 0.5 \text{ rad}, \\
  | \theta | &\leq 0.5 \text{ rad},
  \end{aligned}
  \qquad
  \begin{aligned}
    |\dot{v}_{z,cmd}| &\leq 2.0 \text{ m/s}, \\
  |\phi_{cmd}| &\leq 0.5 \text{ rad}, \\
  |\theta_{cmd}| &\leq 0.5 \text{ rad}, \\
  |\dot{\psi}_{cmd}| &\leq \pi/2 \text{ rad/s}. 
  \end{aligned}
\end{equation*}
The constraints in the left column constitute the set $\mathcal{X}_q$. The first two constraints are related to the maximum velocity and vertical velocity respectively, which we want to limit to prevent fast descent. The latter two are constraints on the roll and pitch angles. The set $\mathcal{U}_q$ is formed of constraints in the right column.

The boat model is chosen as a simple vehicle dynamical model for the purpose of this work. The state vector of boat model $x_b$ is chosen as $x_b = \left [ p_x,\ p_y,\ \psi,\ v_x,\ v_y,\ \omega_\psi \right ]^T,$ and input $u_b = \left [ \tau_x,\ \tau_y,\ \tau_{\omega_\psi} \right ]^T$. The position in $\mathbb{R}^3$ space is represented with $y_b = \left [ p_x,\ p_y,\ 0 \right ]^T$. Matrix $C_b$ is given as $C_b=[\text{diag}(1,1,0),0_{3\times3}]$.

The boat model set constraints $\mathcal{X}_b$ also has the velocity constraints and constraint on the state $\omega_\psi$, i.e. $\sqrt{v_x^2+v_y^2} \leq 15.0 \text{ m/s}$ and $| \omega_\psi | \leq 0.5 \text{ rad/s}.$ Finally, the input constraints $\mathcal{U}_b$ has constraints on $\tau_{\omega_\psi}$, i.e. $| \tau_{\omega_\psi} | \leq 0.5  \text{ rad/s}^2$.

\subsection{Results}
Algorithm~\ref{alg:et_dmpc_rendezvous} is initialized with the following parameters. The planning horizon is set as $T=3$s and sampling period is $\delta=0.1$s for both agents. The update parameters for $\theta(t_k)$ are $\eta=0.1$ and $\varepsilon=0.1$. For the quadcopter we choose the weighting matrices as $Q_q = \text{diag}(30,30,6,1,1,1,1,1,1)$, $R_q=I$ and obtain $P_q$ and $\bar{\alpha}_q=0.2064$ according to Lemma~\ref{lem:terminal}. 
For the boat $Q_b=\text{diag}(5,5,1,1,1,1)$, $R_b=I$, $\bar{\alpha}_b=0.7129$. This choice of the tuning parameters prioritizes the  synchronization of the agent's position in the $xy$-plane such that the quadcopter is above the boat and landing platform before the final descent.

\begin{figure}[ht]
\begin{center}
\includegraphics[width=7.5cm]{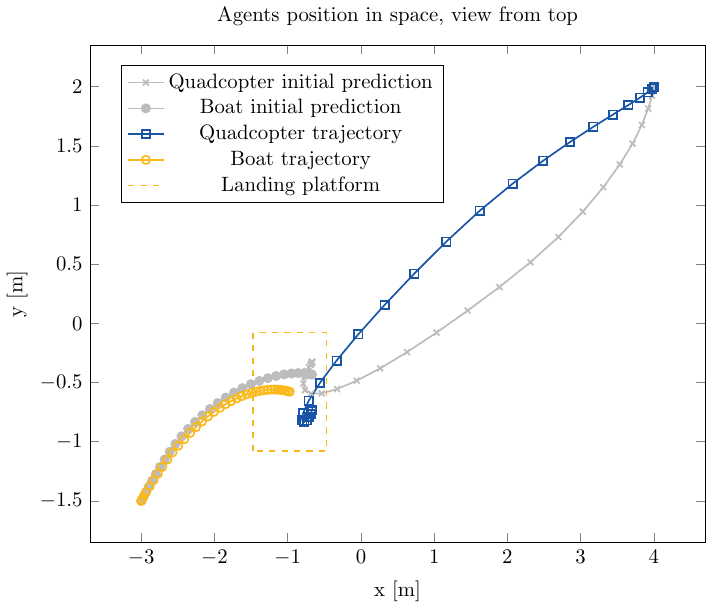}    
\caption{Nominal case with terminal constraints.} 
\label{fig:nominal_without_terminal_top_view}
\end{center}
\end{figure}

\begin{figure}[ht]
\begin{center}
\includegraphics[width=8.8cm]{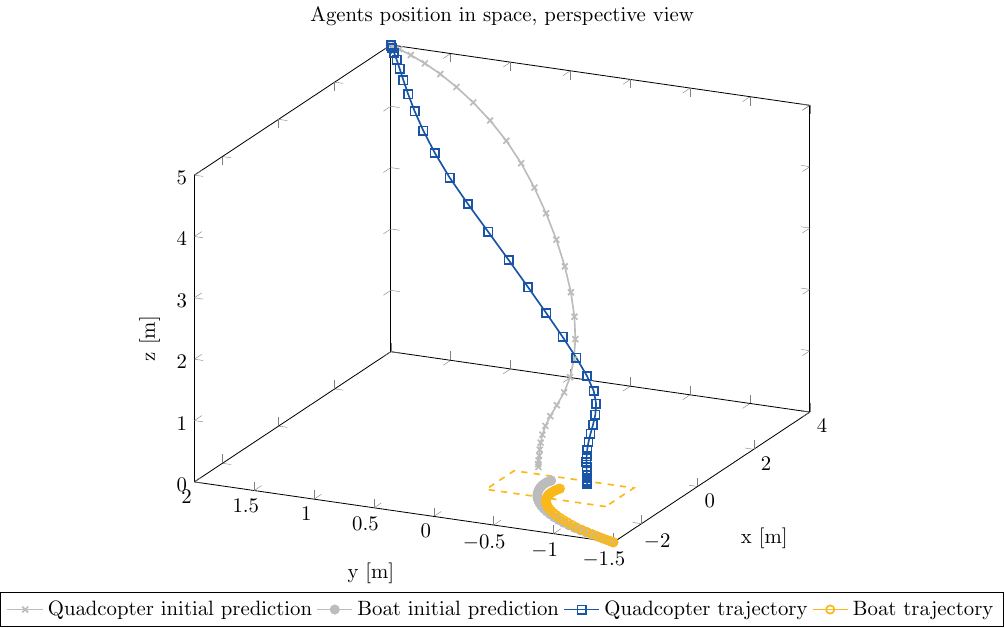}    
\caption{Perspective view of the setup for nominal case with terminal constraints.} 
\label{fig:nominal_without_terminal_perspective}
\end{center}
\end{figure}

We set the initial states of the quadcopter and boat such that the position in the output space is $y_q=[4,2,5]^T$ and $y_b=[-3,-1.5,0]^T$, respectively. To determine initial $\theta(t_0)$ according to Eq.~\eqref{eq:theta_init} we choose $w_q = 2/3$ and $w_b=4/3$. If the initial $\theta(t_0)$ is not changed then the agents will rendezvous at a point $\theta(t_0)=[-0.67,-0.33,0]^T$ that is twice closer to the boat than to the quadcopter as the boat is slower. This is visible in Fig.~\ref{fig:nominal_without_terminal_top_view} for the nominal case with terminal constraints without any disturbances. The difference between the initially predicted and actual trajectories results from the change of $\theta(t_k)$ that occurred for the first four steps and $\theta(t_{final})=[-0.67,-0.73,0]^T$. A perspective view of the same setup is shown in Fig.~\ref{fig:nominal_without_terminal_perspective}.

\begin{figure}[ht]
\begin{center}
\includegraphics[width=8.8cm]{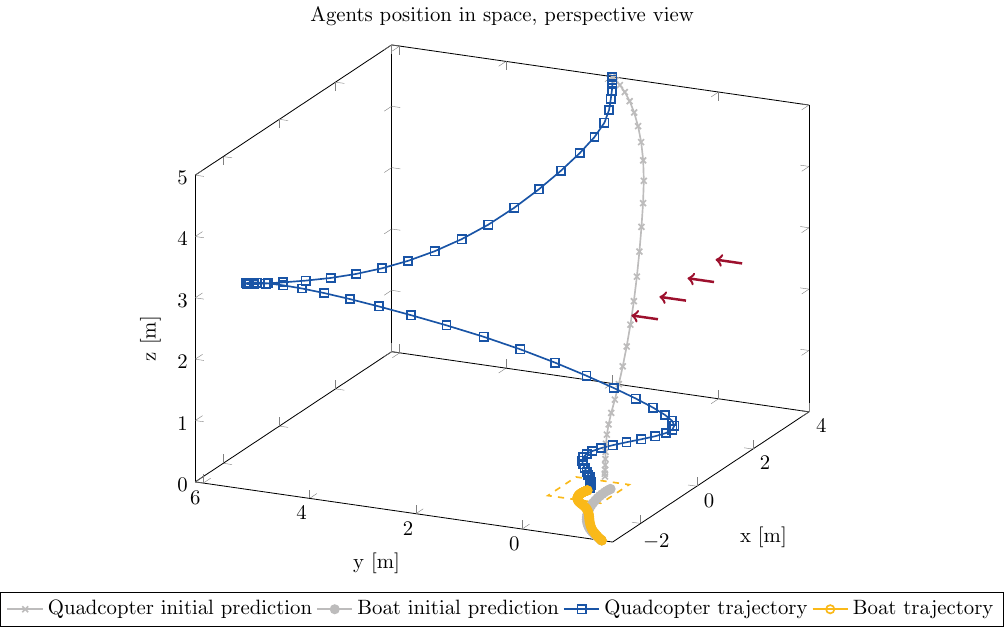}    
\caption{Strong wind active for $t=[0.5\text{s},2.0\text{s}]$, case with terminal constraints. Arrows show wind direction. }
\label{fig:disturbance_with_terminal_perspective}
\end{center}
\end{figure}

In order to show the performance of the update rule for $\theta(t_k)$ we added a strong wind disturbance in the positive $y$-axis direction acting from $t_{1}=0.5$s until $t_{2}=2$s, depicted in Fig.~\ref{fig:disturbance_with_terminal_perspective}. This causes the quadcopter to drift several meters in the direction of the disturbance. However, the feasibility is preserved at all time steps, and because of the imposed terminal constraints the updates of $\theta(t_k)$ are small.

\begin{figure}[ht]
\begin{center}
\includegraphics[width=8.8cm]{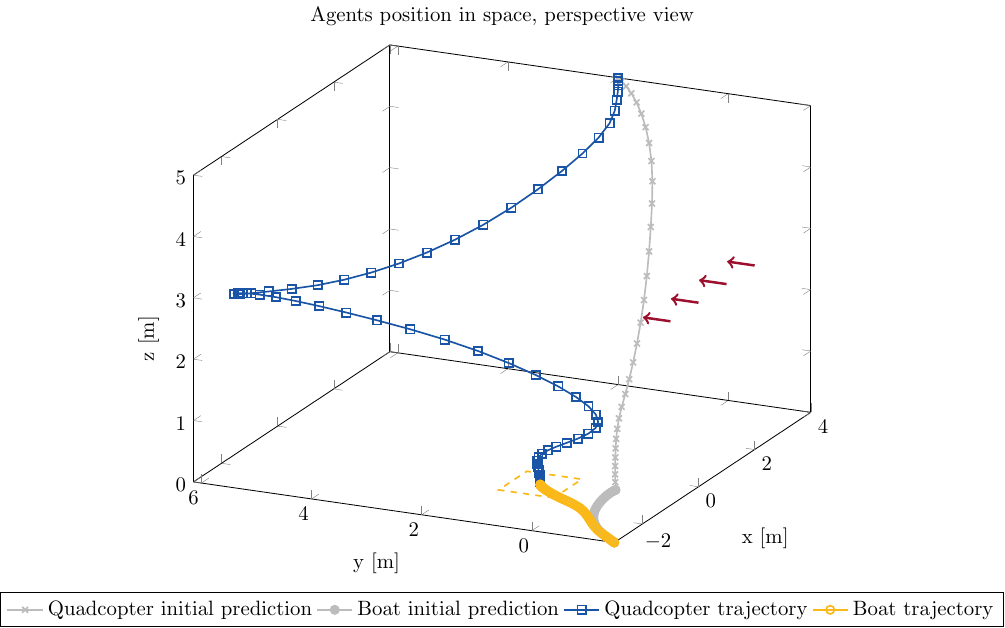}    
\caption{Strong wind active for $t=[0.5\text{s},2.0\text{s}]$, case without terminal constraints. Arrows show wind direction.}
\label{fig:disturbance_without_terminal_perspective}
\end{center}
\end{figure}

\begin{figure}[t]
\begin{center}
\includegraphics[width=7.5cm]{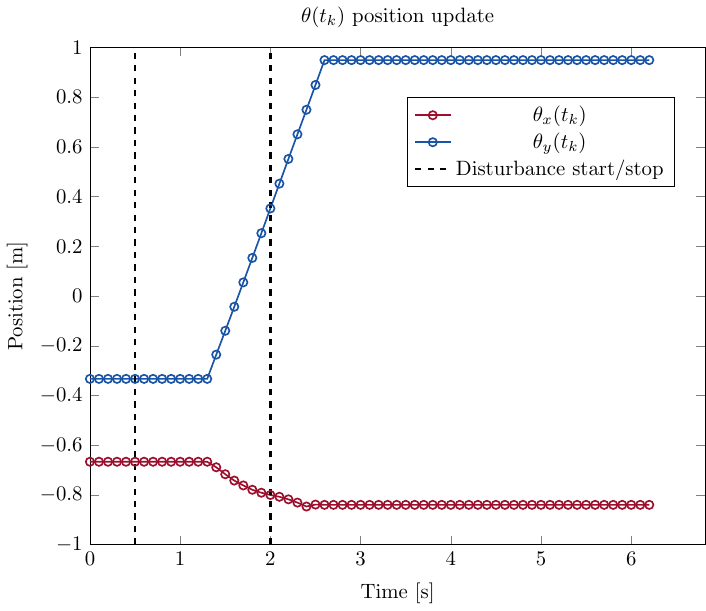}    
\caption{$\theta(t_k)$ evolution in time for the case without terminal constraints}
\label{fig:theta_disturbance_without_terminal_vs_time}
\end{center}
\end{figure}

Finally, because we did not experience any feasibility issues, we removed the terminal constraints from Problem~\ref{prob:docp} to test Algorithm~\ref{alg:et_dmpc_rendezvous} and the update rule. In Fig.~\ref{fig:disturbance_without_terminal_perspective} we can notice that the boat made adjustments and approached to the quadcopter as a result of the rendezvous point updates by the quadcopter. The updates of $\theta(t_k)$ are shown in Fig.~\ref{fig:theta_disturbance_without_terminal_vs_time}. The bigger changes in $\theta(t_k)$ compared to the case with the terminal constraints are due to the update rule. $V_o(\hat{y}_i(t_k+T;t_k),\theta(t_k))$ is evaluated at the last predicted $\hat{y}_i(t_k+T;t_k)$ output for which the corresponding state $\hat{x}_i(t_k+T;t_k)$ belongs to a very small set $\mathcal{X}_{f,i}(\bar{x}_i,\bar{\alpha}_i)$. 

\section{Conclusion}

In this paper, we presented a rendezvous algorithm for the distributed MPC scheme for agents with nonlinear and heterogeneous dynamics. The algorithm is designed for the problem of autonomous cooperative landing of the quadcopter on the autonomous boat. During the landing the agents communicate only when it is necessary to update the rendezvous point and ensure the feasibility of the algorithm. The effectiveness of the proposed algorithm is shown with the simulation of the landing scenarios.

Although we did not experience feasibility issues, in future work, we aim to quantify the upper bound on the disturbance such that the feasibility of the algorithm is preserved.
Furthermore, it will be interesting to include the obstacles and constraints in the output space and examine the behaviour of the algorithm on the real systems.  

\bibliography{ifacconf}             

\begin{thebibliography}{17}
\providecommand{\natexlab}[1]{#1}
\providecommand{\url}[1]{\texttt{#1}}
\providecommand{\urlprefix}{URL }
\expandafter\ifx\csname urlstyle\endcsname\relax
  \providecommand{\doi}[1]{doi:\discretionary{}{}{}#1}\else
  \providecommand{\doi}{doi:\discretionary{}{}{}\begingroup
  \urlstyle{rm}\Url}\fi

\bibitem[{Bereza et~al.(2020)Bereza, Persson, and Wahlberg}]{berezapersson}
Bereza, R., Persson, L., and Wahlberg, B. (2020).
\newblock Distributed model predictive control for cooperative landing.
\newblock In \emph{2020 Proceedings of IFAC World Congress}. IFAC.

\bibitem[{Chen and Allg{\"o}wer(1998)}]{chen1998quasi}
Chen, H. and Allg{\"o}wer, F. (1998).
\newblock A quasi-infinite horizon nonlinear model predictive control scheme
  with guaranteed stability.
\newblock \emph{Automatica}, 34(10), 1205--1217.

\bibitem[{Christofides et~al.(2013)Christofides, Scattolini, de~la Pena, and
  Liu}]{christofides2013distributed}
Christofides, P.D., Scattolini, R., de~la Pena, D.M., and Liu, J. (2013).
\newblock Distributed model predictive control: A tutorial review and future
  research directions.
\newblock \emph{Computers \& Chemical Engineering}, 51, 21--41.

\bibitem[{Dunbar(2007)}]{dunbar2007distributed}
Dunbar, W.B. (2007).
\newblock Distributed receding horizon control of dynamically coupled nonlinear
  systems.
\newblock \emph{IEEE Transactions on Automatic Control}, 52(7), 1249--1263.

\bibitem[{Dunbar and Murray(2006)}]{dunbar2006distributed}
Dunbar, W.B. and Murray, R.M. (2006).
\newblock Distributed receding horizon control for multi-vehicle formation
  stabilization.
\newblock \emph{Automatica}, 42(4), 549--558.

\bibitem[{Hashimoto et~al.(2014)Hashimoto, Adachi, and
  Dimarogonas}]{hashimoto2014distributed}
Hashimoto, K., Adachi, S., and Dimarogonas, D.V. (2014).
\newblock Distributed aperiodic model predictive control for multi-agent
  systems.
\newblock \emph{IET Control Theory \& Applications}, 9(1), 10--20.

\bibitem[{Hashimoto et~al.(2017)Hashimoto, Adachi, and
  Dimarogonas}]{hashimoto2017event}
Hashimoto, K., Adachi, S., and Dimarogonas, D.V. (2017).
\newblock Event-triggered intermittent sampling for nonlinear model predictive
  control.
\newblock \emph{Automatica}, 81, 148--155.

\bibitem[{Heemels et~al.(2012)Heemels, Johansson, and
  Tabuada}]{heemels2012introduction}
Heemels, W., Johansson, K.H., and Tabuada, P. (2012).
\newblock An introduction to event-triggered and self-triggered control.
\newblock In \emph{2012 IEEE 51st IEEE Conference on Decision and Control
  (CDC)}, 3270--3285. IEEE.

\bibitem[{Kamel et~al.(2017)Kamel, Burri, and Siegwart}]{kamel2017linear}
Kamel, M., Burri, M., and Siegwart, R. (2017).
\newblock Linear vs nonlinear mpc for trajectory tracking applied to rotary
  wing micro aerial vehicles.
\newblock \emph{IFAC-PapersOnLine}, 50(1), 3463--3469.

\bibitem[{Keviczky et~al.(2006)Keviczky, Borrelli, and
  Balas}]{keviczky2006decentralized}
Keviczky, T., Borrelli, F., and Balas, G.J. (2006).
\newblock Decentralized receding horizon control for large scale dynamically
  decoupled systems.
\newblock \emph{Automatica}, 42(12), 2105--2115.

\bibitem[{Keviczky and Johansson(2008)}]{keviczky2008study}
Keviczky, T. and Johansson, K.H. (2008).
\newblock A study on distributed model predictive consensus.
\newblock \emph{IFAC Proceedings Volumes}, 41(2), 1516--1521.

\bibitem[{Lim{\'o}n et~al.(2008)Lim{\'o}n, Alvarado, Alamo, and
  Camacho}]{limon2008mpc}
Lim{\'o}n, D., Alvarado, I., Alamo, T., and Camacho, E.F. (2008).
\newblock Mpc for tracking piecewise constant references for constrained linear
  systems.
\newblock \emph{Automatica}, 44(9), 2382--2387.

\bibitem[{Liu et~al.(2020)Liu, Li, Shi, and Xu}]{liu2020distributed}
Liu, C., Li, H., Shi, Y., and Xu, D. (2020).
\newblock Distributed event-triggered model predictive control of coupled
  nonlinear systems.
\newblock \emph{SIAM Journal on Control and Optimization}, 58(2), 714--734.

\bibitem[{M{\"u}ller et~al.(2012)M{\"u}ller, Reble, and
  Allg{\"o}wer}]{muller2012cooperative}
M{\"u}ller, M.A., Reble, M., and Allg{\"o}wer, F. (2012).
\newblock Cooperative control of dynamically decoupled systems via distributed
  model predictive control.
\newblock \emph{International Journal of Robust and Nonlinear Control}, 22(12),
  1376--1397.

\bibitem[{Persson and Wahlberg(2019)}]{persson2019model}
Persson, L. and Wahlberg, B. (2019).
\newblock Model predictive control for autonomous ship landing in a search and
  rescue scenario.
\newblock In \emph{AIAA Scitech 2019 Forum}, 1169.

\bibitem[{Persson and Wahlberg(2021)}]{perssonBW}
Persson, L. and Wahlberg, B. (2021).
\newblock Variable prediction horizon control for cooperative landing on moving
  target.
\newblock In \emph{2021 IEEE Aerospace Conference}. IEEE.

\bibitem[{Richards and How(2007)}]{richards2007robust}
Richards, A. and How, J.P. (2007).
\newblock Robust distributed model predictive control.
\newblock \emph{International Journal of control}, 80(9), 1517--1531.

\end{thebibliography}
                                                   







\appendix

 \section{Proof of Lemma~\ref{lem:moving_terminal}} \label{ap:appA}
\begin{pf}
Let us consider the optimal control law $\hat{u}_i^*(s;t_k)$ for interval $s \in \left [ t_k,t_k+T \right ]$ obtained at $t_k$ by solving Problem~\ref{prob:docp} and a candidate control law 
\begin{equation}\label{eq:control_candidate}
    \tilde{u}_i(s;t_{k+1}) = \left\{\begin{matrix}
\hat{u}_i^*(s;t_k) & s\in \left [ t_{k+1},t_k+T \right ] \\
K_i\tilde{x}_i(s;t_k) & s \in \left [ t_k+T,t_{k+1}+T \right ] 
\end{matrix}\right.\end{equation} 
that generates the system trajectory $\tilde{x}_i(s;t_{k+1})$ based on the dynamics in \eqref{eq:nominal}. It holds that $\tilde{x}_i(t_k+T;t_{k+1}) \in \mathcal{X}_{f,i}(\bar{x}_i(t_k),\alpha_i(t_k))$ and, due to the invariance of the terminal set, $\tilde{x}_i(t_{k+1}+T;t_{k+1}) \in \mathcal{X}_{f,i}(\bar{x}_i(t_k),\alpha_i(t_k))$, i.e.
\begin{equation}\label{eq:alpha_i_def}
    \left \| \tilde{x}_i(t_{k+1}+T;t_{k+1}) - \bar{x}_i(t_k) \right \|_{P_i}^2 \leq \alpha_i^2(t_k).
\end{equation}
Then, 
\begin{align*}
\begin{split}
     &\left \| \tilde{x}_i(t_{k+1}+T;t_{k+1}) - \bar{x}_i(t_{k+1}) \right \|_{P_i} \\
     &\leq \left \| \tilde{x}_i(t_{k+1}+T;t_{k+1}) - \bar{x}_i(t_k) \right \|_{P_i} + \left \| \bar{x}_i(t_k)- \bar{x}_i(t_{k+1}) \right \|_{P_i} \\
     &\overset{\eqref{eq:alpha_i_def}}{\leq}\alpha_i(t_k) + \left \| \bar{x}_i(t_k)- \bar{x}_i(t_{k+1}) \right \|_{P_i} \\
     &\overset{\eqref{eq:steady_states_alpha}}{=} \alpha_i(t_k) + \left \| H_{x_i}\theta_i(t_k)- H_{x_i}\theta_i(t_{k+1}) \right \|_{P_i}   \\
     &\overset{\eqref{eq:theta_update}}{=}\alpha_i(t_k) + \eta \left \| H_{x_i}v_\theta(t_k) \right \|_{P_i} = \alpha_i(t_{k+1}).
     \end{split}
\end{align*}
Hence, $\tilde{x}_i(t_{k+1}+T;t_{k+1}) \in \mathcal{X}_{f,i}(\bar{x}_i(t_{k+1}),\alpha_i(t_{k+1}))$.
\end{pf}
\section{Proof of Theorem~\ref{thm:recursive_feasibility}}\label{ap:appB}
\begin{pf}
If the state $x_i(t_{k+1})\in \mathcal{X}_{f,i} \subseteq \mathcal{X}_i$ then by the invariance of the terminal set stated in Lemma \ref{lem:terminal}, it will remain in that set. Therefore, using the terminal control law $\kappa_{f_i}(x)=K_i x_i \in \mathcal{U}_i$, the cost function in \eqref{eq:cost} is bounded and all constraints in \eqref{eq:prob} are satisfied.

Let us consider again the obtained optimal control law $\hat{u}_i^*(s;t_k)$ at $t_k$ for interval $s \in \left [ t_k,t_k+T \right ]$ and a candidate control law according to Eq. \eqref{eq:control_candidate}
that generates the system trajectory $\tilde{x}_i(s;t_{k+1})$ based on the dynamics in  \eqref{eq:nominal}.

Because of feasibility at $t_k$, the state $\tilde{x}_i(s;t_{k+1}) \in \mathcal{X}_i$ for $s \in \left [ t_{k+1},t_k+T \right ]$ and $\tilde{x}_i(t_k+T;t_{k+1})\in \mathcal{X}_{f,i}(\bar{x}_i(t_k),\alpha_i(t_k))$. Moreover, due to the terminal set properties from Lemma \ref{lem:terminal}, and the result of Lemma \ref{lem:moving_terminal} the candidate control law will ensure that the terminal state $\tilde{x}_i(t_{k+1}+T;t_{k+1})$ is in the shifted local terminal set $\tilde{x}_i(t_{k+1}+T;t_{k+1})\in \mathcal{X}_{f,i}(\bar{x}_i(t_{k+1}),\alpha_i(t_{k+1}))$, which proves recursive feasibility. 
\end{pf}

\end{document}